\documentclass[aps,pre,twocolumn,10pt,amsmath,amsfonts,amssymb,floatfix,bibnotes]{revtex4}
\usepackage{amssymb}
\usepackage{graphicx,color}
\usepackage{bbm}
\usepackage{multirow}

\usepackage{amsmath,bm,epsfig}
\newcommand{\B}[1]{{\bm{#1}}}

\begin{document}

\title{Training, Memory and Universal Scaling in Amorphous Frictional Granular Matter}

\author{M. M. Bandi$^1$}
\author{H. George E. Hentschel$^2$}
\author{ Itamar Procaccia$^{2,3}$}
\author{Saikat Roy$^2$}
\author{Jacques Zylberg$^2$}
\affiliation{$^1$ Collective Interactions Unit, OIST Graduate University, Onna, Okinawa, 904-0495 Japan. \\
$^2$Dept of Chemical Physics, The Weizmann Institute of Science, Rehovot 76100, Israel.\\$^3$The Niels Bohr International Academy, University of Copenhagen, Blegdamsvej 17, DK-2100 Copenhagen, Denmark}

\begin{abstract}
We report a joint experimental and theoretical investigation of cyclic training of amorphous frictional granular assemblies, with special attention to memory formation and retention. Measures of dissipation and compactification are introduced, culminating with a proposed scaling law
for the reducing dissipation and increasing memory. This scaling law is expected to be universal, insensitive to the details of the elastic and frictional interactions between the granules.
\end{abstract}
\maketitle

``Memory" in materials physics is usually associated with the existence of macroscopic hysteretic responses \cite{08Lag}. Two distinct states, separated by a potential barrier larger than the thermal energy scale, can be used as a memory encoding mechanism.
Here we focus
on memory that is induced by training a frictional granular matter by cyclic loading and
unloading \cite{08CCGP,11KN,14PKN,14KA,13BRKE,13RLR,14FFS, 17AB}.  In each such cycle dissipation leads to hysteresis, but with repeated cycles the dissipation diminishes until the system retains memory of an asymptotic loaded state that is not forgotten even under complete unloading. We report and explain a universal power law associated with the reduced dissipation and increase in memory which is expected to hold irrespective of the details of the microscopic interactions.
\begin{figure}[h!]
\includegraphics[scale=0.30]{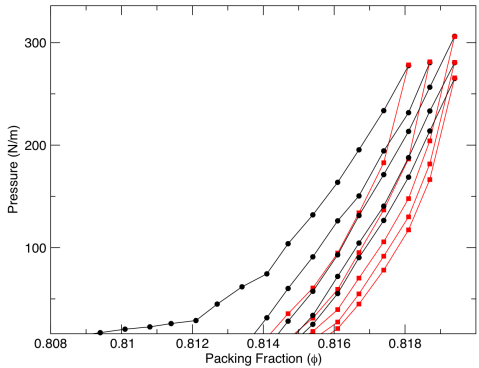}
\caption{Typical hysteresis loops obtained experimentally upon uniaxial compression and decompression
of an amorphous configuration of frictional disks as seen in Fig.~\ref{incon}. The pressure $P$ is measured in $N/m$, and $\Phi$ is dimensionless. Compression
legs are in black and decompression in red.}
\label{exploops}
\end{figure}

The phenomenon under study is best introduced by the experimental plots of pressure vs. packing fraction obtained by compressing and decompressing uniaxially an array of frictional disks \cite{13BRKE,17AB}, cf. Fig.~\ref{exploops}. The experimental set up is detailed in the SI. A typical example of the experimental cell is shown in the upper panel of Fig.~\ref{incon}. Two opposing boundaries separated by chamber length L were movable while the other two (transverse) boundaries were held fixed. The two opposing movable boundaries provided uni-axial pack compression, through which the packing fraction $\Phi$ was controlled. Accordingly, we define the packing fraction $\Phi$ as the ratio of total area occupied by the disks to the chamber area bounded within the four boundaries, two of which are movable. Fig.~\ref{exploops} displays a typical series of consecutive loops of compression and decompression. The qualitative experimental observation is that the area in consecutive hysteresis loops diminishes monotonically while the packing fraction is increasing with every loop. This indicates that the system is compacted further with every loop and this process is accompanied by a reduction in the dissipation. The experimental results left however two open questions: (i) whether
asymptotically the dissipation vanished, such that every compression became purely elastic
and the decompression to zero pressure left the system with perfect memory of the stressed
configuration; and (ii) whether there is anything universal in the way that the areas of the loops
approaches its asymptote, be them finite or zero. To answer these question we performed numerical simulations that lead to the conclusion that (i) asymptotically the hysteresis loops are still dissipative due to frictional losses, but the
structural rearrangements disappear and the neighbor list becomes invariant; and (ii)
that the area $A_n$ under the $n$th hysteresis loop
(which is a direct measure of the dissipation) decays as a power law to an asymptotic value according to
\begin{equation}
A_n = A_\infty +B n^{-\theta} \ , \quad \theta\approx 1 \label{plaw} \ .
\end{equation}
Here $A_\infty$ represents the dissipation due to frictional slips that exist even in the asymptotic loop, and it depends on the material properties. The second
term in Eq.~(\ref{plaw}) is due to the successive compactification of the sample, and the constant $B$ is also expected to depend on material properties. The form of this law however is universal, expected to hold independently of the details of the microscopic interaction between the granules.

The details of the numerical set up are provided in the SI.  An example of an initial
\begin{figure}
\includegraphics[scale=0.20]{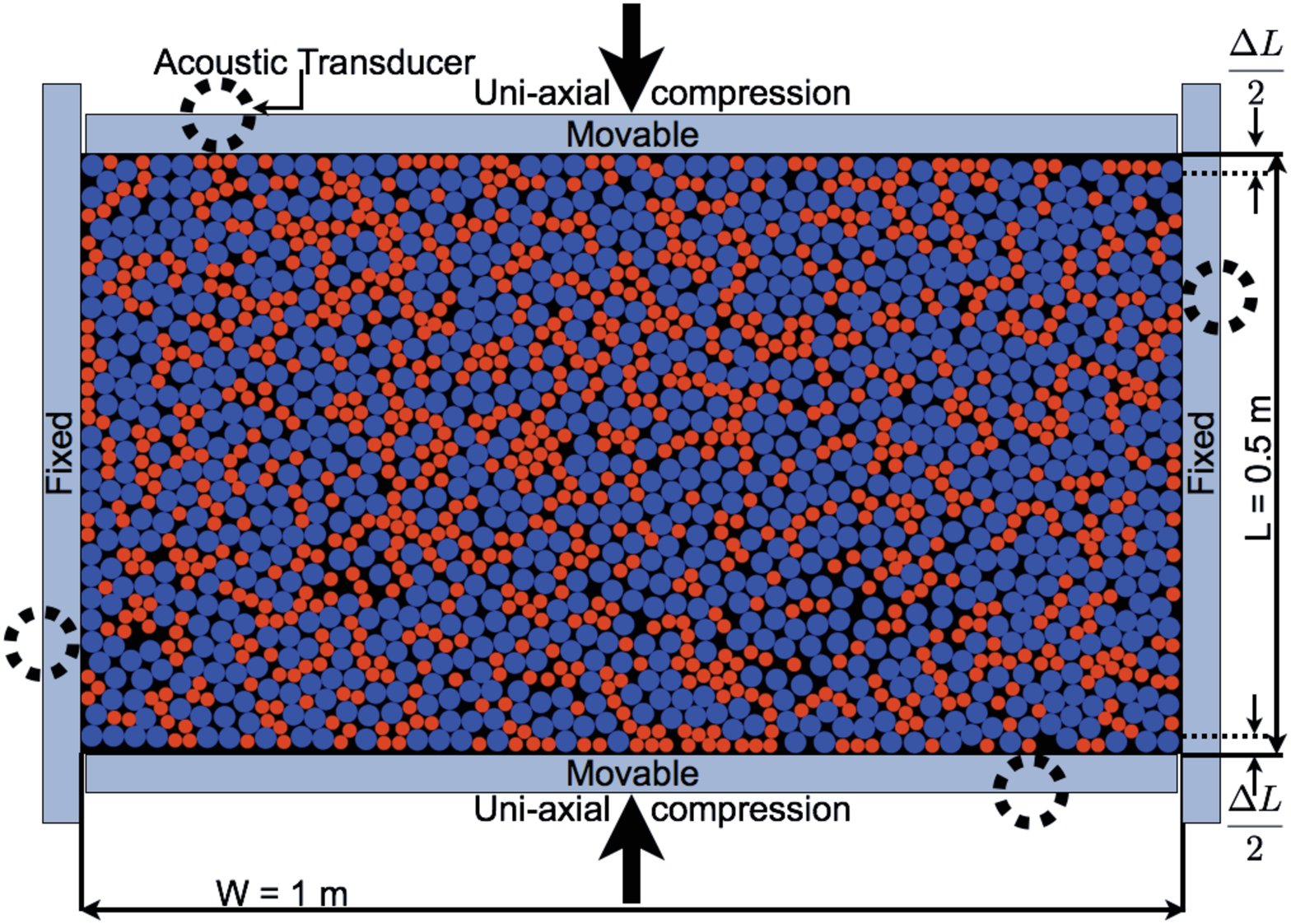}
\vskip 0.1 cm
\includegraphics[scale=0.35]{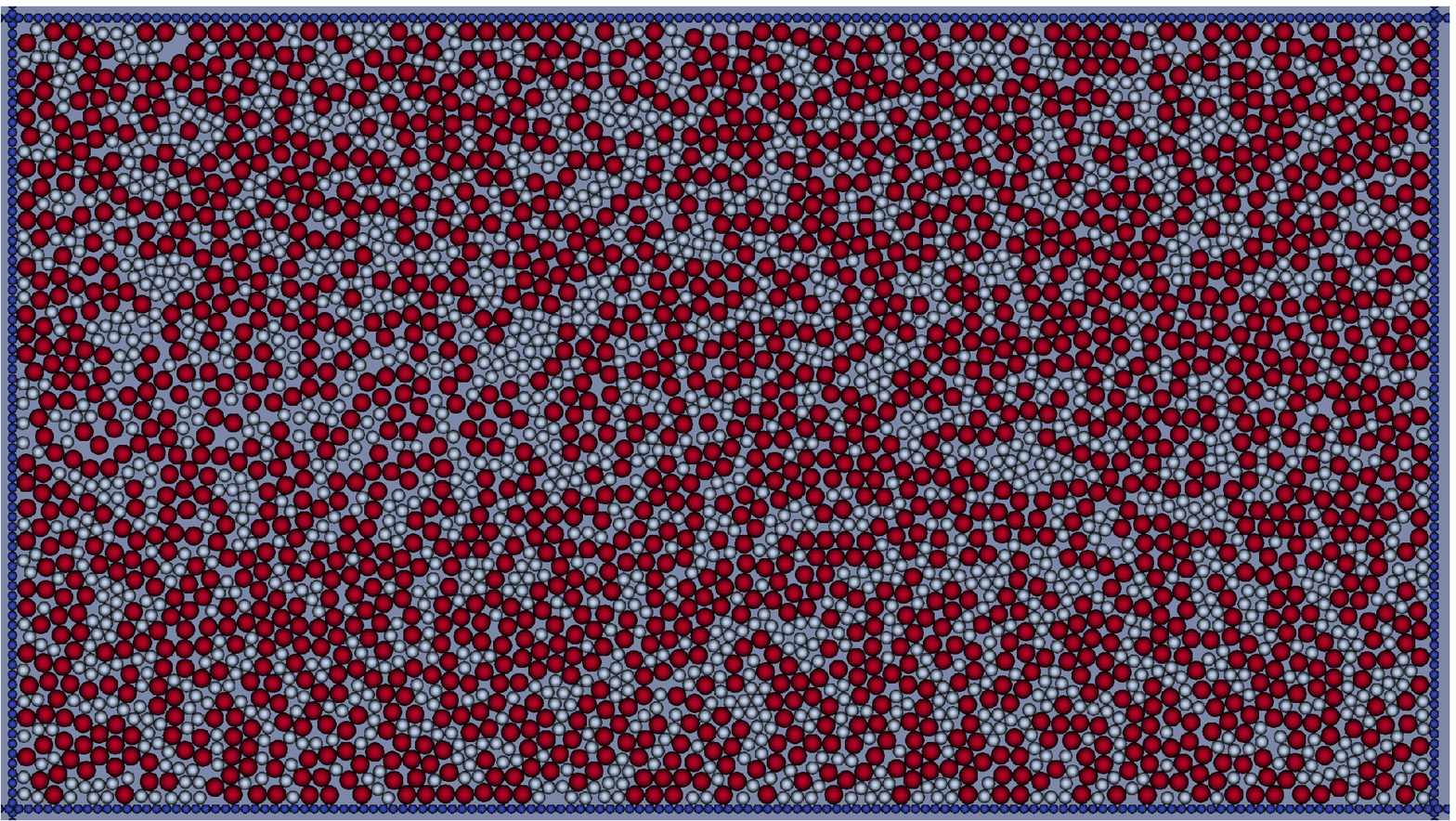}
\caption{Upper panel: an example of a typical initial configuration in the experiment. Lower pannel: an example
of a typical initial configuration in the numerical simulation.}
\label{incon}
\end{figure}
configuration is shown in the lower panel of Fig.~\ref{incon}. We assign Hertzian normal force $F^{(n)}_{ij}$ and a Mindlin tangential force $F^{(t)}_{ij}$ \cite{79CS} to each binary contact $ij$. The tangential force is always limited
by the Coulomb law
   \begin{equation}
   F^{(t)}_{ij} \le \mu F^{(n)}_{ij} \ . \label{Coulomb}
   \end{equation}
In uniaxial straining the pressure is increased by pushing two opposite walls of the system towards each other. In each cycle we first reach a chosen maximal pressure by quasi-static steps. After each compression step, the system is allowed to relax to reach a new mechanical equilibrium in which the global stress tensor is measured by averaging the dyadic products between all the binary contact forces and the vectors connecting the centers of mass in a given volume. The trace of this stress tensor is the new pressure $P$. After a full compression leg, a cycle is completed by decompressing back to zero pressure, where the next compression cycle begins. The packing fraction $\Phi$ is monitored throughout this process. Each such cycle traces a hysteresis loop in the $P-\Phi$ plane, see Fig.~\ref{loops} as an example. The area within each hysteresis loop is a measure of the dissipation, which
\begin{figure}
\includegraphics[scale=0.22]{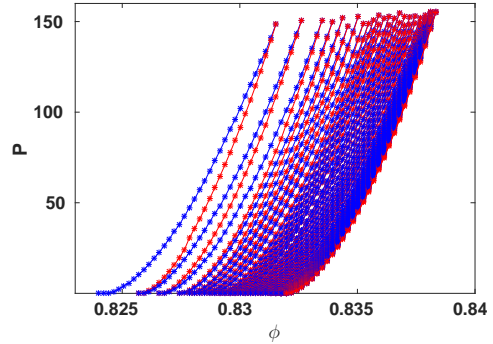}
\caption{A succession of hysteresis loops as measured in the numerical simulation. Blue symbols are compression legs and red symbols decompression legs. Here $\mu=0.1$.}
\label{loops}
\end{figure}
in general stems from two sources. One is plastic events in which the neighbor lists change in
an irreversible fashion, and the other is due to frictional losses when the frictional
tangential force exceeds the allowed Coulomb limit. The training of the system is exemplified
by the fact that the dissipation as measured by the area $A_n$ of the $n$th cycle reduces with
$n$ and reaches an asymptotic value when $n\to \infty$. Measuring the area in the $n$th loop we
find that it follows a power law decay in the form of Eq.~(\ref{plaw})
The data supporting this power law are exhibited in Fig.~\ref{power}.
\begin{figure}
\includegraphics[scale=0.20]{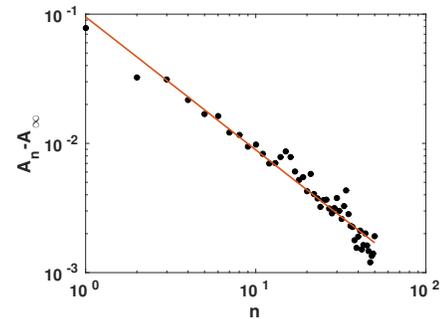}
\caption{The power law for the decaying areas under the hysteresis loops as measured in the numerical simulation. Here $\mu=0.1$,  Black dots are data and the red line is the best fitting power law $y=0.095 X^{-1.03}$. The observed lot is independent of $\mu$, cf. the SI.}
\label{power}
\end{figure}
From this data we can conclude that $\theta\approx 1$ and that the scaling law appears
universal with respect to changes in the value of $\mu$. A further evidence of universality
is obtained by changing the size distribution of disks, choosing a multi-dispersed system
with readii ratios 1, 1.1, 1.2 and 1.4. Identical power laws were found.

To understand the scaling law we need to identify the two important processes that take place during
the cyclic training. One is compactification. In every compression leg of the cycle the system compactifies, until a limit $\phi$ value is reached for the chosen maximal pressure.  To quantify this process we can measure the volume fraction $\Phi_n(P_{\rm max})$ at the highest value of the pressure in the $n$th cycle. Define then a new variable
\begin{equation}
X_n \equiv \Phi_{n+1}(P_{\rm max})-\Phi_n(P_{\rm max}) \ .
\end{equation}
This new variable is history dependent in the sense that $X_{n+1} = g(X_n)$ where the function $g(x)$ is unknown at this
point. This function must have a fixed point $g(x=0) =0$ since the series $\sum_n X_n$ must converge; for any given
chosen maximal pressure there is a limit volume fraction that cannot be exceeded. Near the fixed point, assuming
analyticity, we must have the form
\begin{equation}
X_{n+1} = g(X_n) = X_n -C X_n^2 + \cdots \ .
\end{equation}
The solution of this equation for $n$ large is
\begin{equation}
X_n = \frac{C^{-1}}{n} \ . \label{scaling}
\end{equation}
This is the source of the second term in Eq.~(\ref{plaw}), which stems from the compactification which reduces the
amount of dissipation due to irreversible plastic rearrangements. A direct measurement of $X_n$ as a function of $n$ is
shown in the log-log plot presented in Fig.~\ref{Xnvsn}, supporting the generality of this power law. Without any reason
for non-analyticity in the function $g(X_n)$ this conclusion is firm.
\begin{figure}
\includegraphics[scale=0.25]{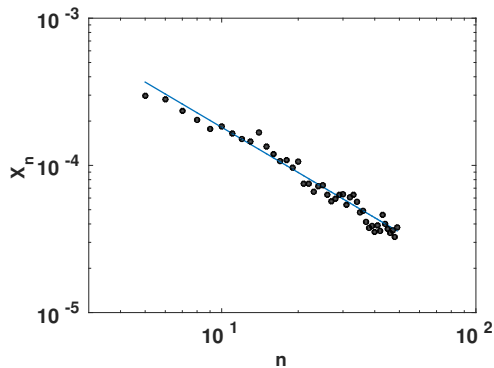}
\caption{Log-log plot of $X_n$ vs. $n$. The black dots are the data, the blue line is the best fitting scaling law
$y=0.02 x^{-1.02}$. The data corroborates Eq.~(\ref{scaling}).}
\label{Xnvsn}
\end{figure}
It should be noted at this point that the scaling laws Eqs.~(\ref{plaw}) and (\ref{scaling}) must contain some logarithmic corrections, since
the harmonic series does not converge, but the series $\sum_n X_n$ must converge to get an asymptotic value of $\Phi_{\rm max}$. Indeed, the scaling laws measured above consistently show exponents slightly smaller than -1, and therefore
the series of $\Phi_n$ converges. It is very likely that this small difference is due to logarithmic corrections to the scaling law which cannot be computed from the simple theory presented here.

The first term in Eq.~(\ref{plaw}), on the
other hand, is due to the frictional dissipation. In uniaxial compression there is a shear component, and the shear stress
loads the tangential contacts. Whenever the tangential force exceeds the Coulomb limit Eq.~(\ref{Coulomb}), the system
dissipates some energy to a frictional slip. Even if the neighbor list becomes invariant at large values of $n$, the loading
of the system is always accompanied by frictional slips. In our simulations we can measure the energy dissipated by friction slips, denoted as  $\Delta E^{(f)}_n$ in the $n$th hysteresis loop. To have a non-dimensional measure we normalized this quantity by $\Delta E^{(f)}_1$, denoting the result as $S_n$. The dependence of this normalized dissipated energy on $n$ is presented
in Fig.~\ref{slips}. It is clear that the normalized dissipated energy due to frictional slips reduces
rapidly to a stable value; this is the first term in Eq.~(\ref{plaw}).
\begin{figure}
\includegraphics[scale=0.22]{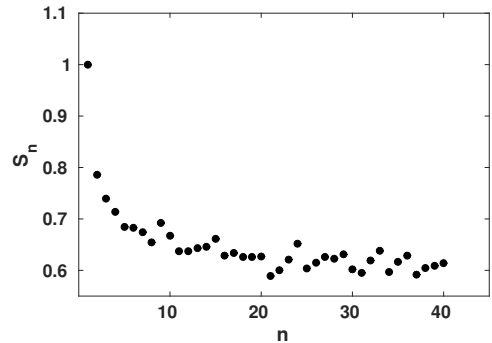}
\caption{The normalized frictional energy loss in each hysteresis loop. This energy loss drops to a stable value
that is responsible for the asymptotic dissipation that is encoded by the area $A_\infty$ in Eq.(\ref{plaw}). }
\label{slips}
\end{figure}

Encouraged by this theory we returned to the experimental data to measure both $X_n$ and $A_n$. To obtain the logarithmic $n$ dependence of the area  we needed to know the value of $A_\infty$. A direct measurement of this area is unfeasible. But one recognizes that the total dissipation under the asymptotic loop should be the same with or without memory. Accordingly, we applied an acoustic perturbation to the configuration after each quasi-static step to destroy memory and training, and force the system to go to asymptotic state. The result is shown in the upper panel of Fig.~\ref{expts}, showing  $A_{\infty}$ alone in blue squares. Indeed  $A_{\infty}$ is finite and flat as a function of $n$. Using the measured value of $A_{\infty}$ we get the power low scaling for $A_n - A_{\infty}$ as expected and as shown by the data with black rhombi. To underline the fact
 that this scaling law is governed by the memory during the training protocol we demonstrate the change that is caused by destroying the memory midway through the cycles; this is a further validation that the power law is indeed coming from training and memory formation. For the data shown in green triangles, there are no acoustic perturbations for $n$ = 1 - 10; there we see the power-law behavior. From $n$ = 11 - 50, we apply acoustic perturbations after each quasi-static step.  The loop areas now fall drastically and become flat. Note that the magnitude is below $A_{\infty}$  since we are plotting the difference $A_n-A_\infty$.
\begin{figure}
\includegraphics[scale=0.25]{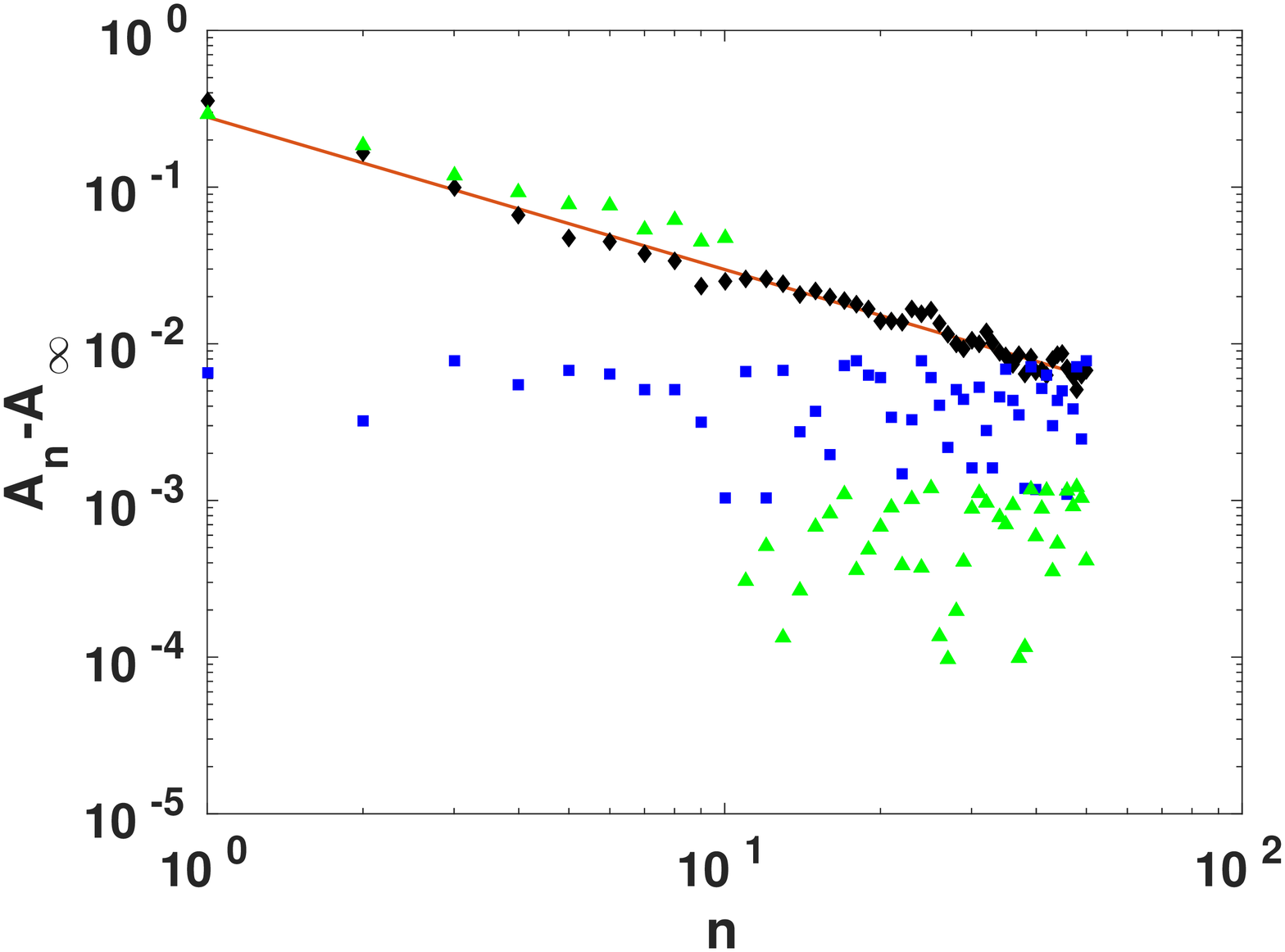}
\includegraphics[scale=0.25]{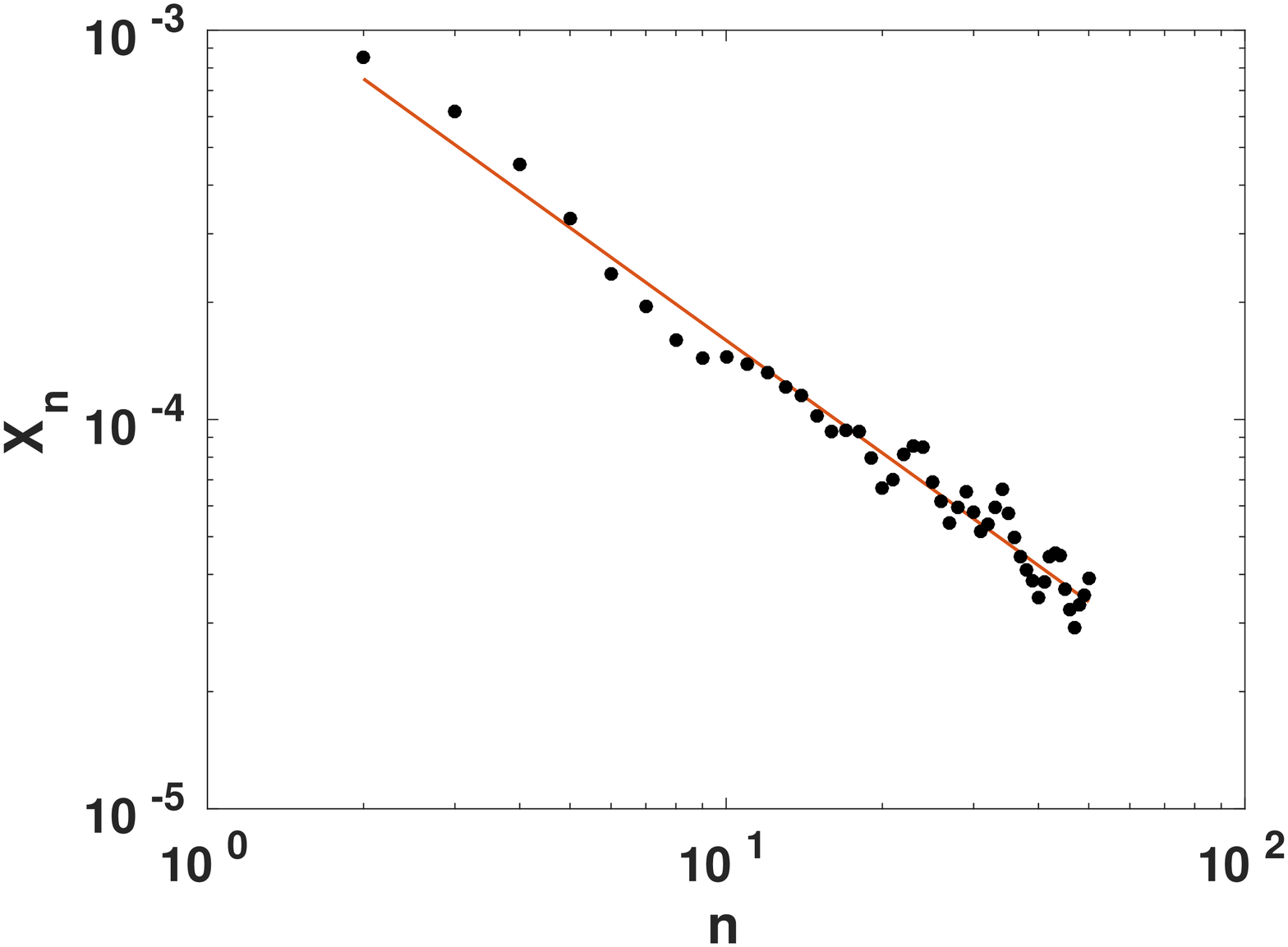}
\caption{Upper panel: The areas $A_n-A_\infty$ measured experimentally as a function of $n$, agreeing
to Eq.~(\ref{plaw}) with $\theta\approx 0.97$; black rhombi. Blue squares: the estimate of $A_\infty$ obtained by forcing
the system to the asymptote by acoustic perturbations. Green triangles: the areas resulting
from the destruction of memory after 10 regular loops. Lower panel: experimental results for $X_n$
shown in log-log plot vs. $n$; the slope is approximately 0.96. }
\label{expts}
\end{figure}

 It should be commented that the simple scenario discussed in the Letter {\em requires} a subtle change in the shape of the hysteresis loops. The low order loops are increasing the volume fraction, such that the compression leg starts with at a lower value of $\Phi$ than the end of the decompression loop, see the upper panel in Fig.~\ref{examples}.
This continues to be the case as long as the systems can be compactified further. The high order loops must begin
and end at the same value of $\Phi_n$, see the lower panel in Fig.~\ref{examples}. Thus for large value of $n$ the hysteresis loops become repetitive, with an invariant trace in the $P-\Phi$ plane, even though they have frictional
dissipation in the limit $n\to \infty$. This subtle change in the shape of the hysteresis loops allows the function
$g(x)$ to have the fixed point at $x=0$ around which the analytic expansion dictates the universality of the power law.

\begin{figure}
\includegraphics[scale=0.20]{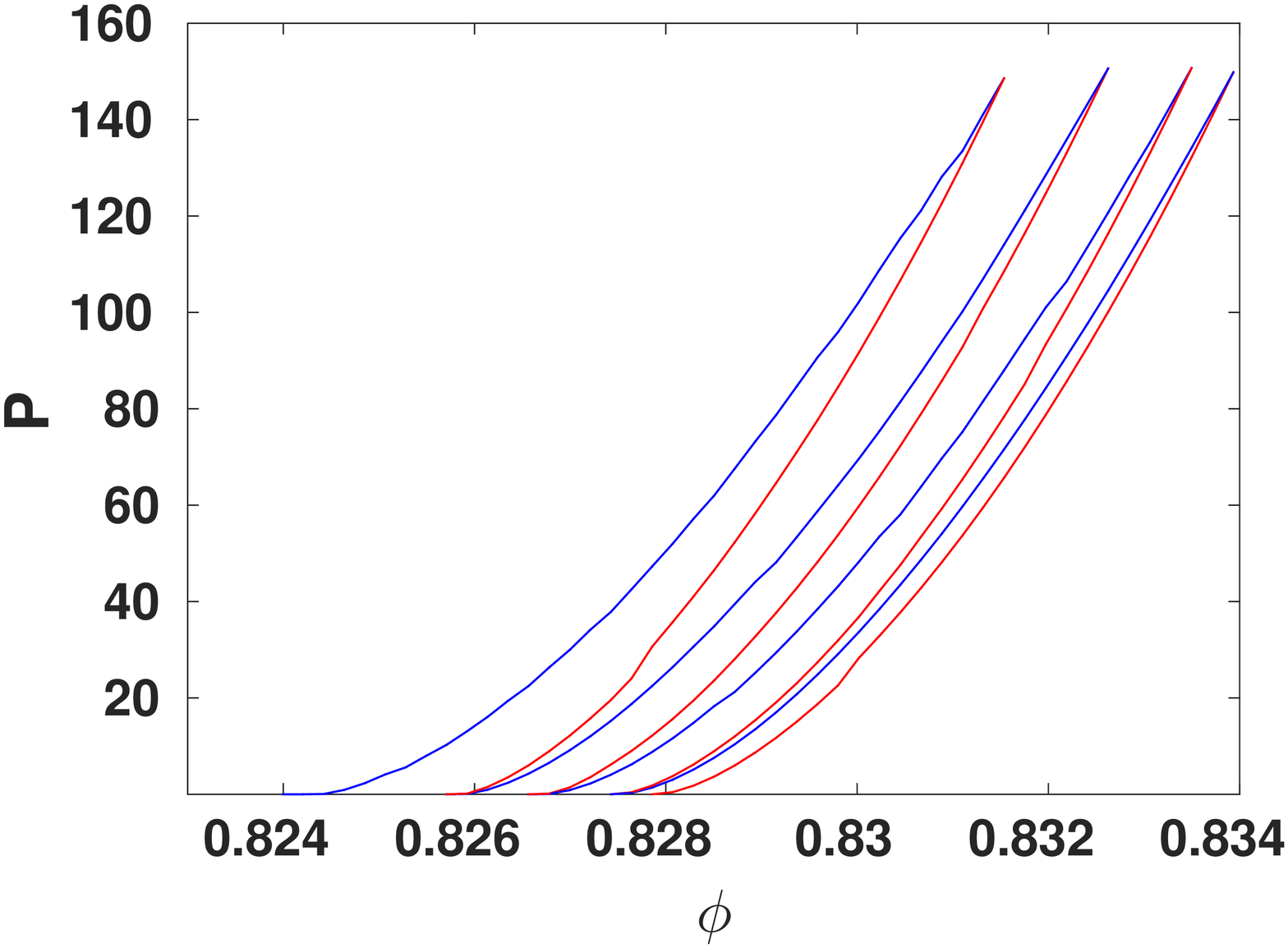}
\includegraphics[scale=0.20]{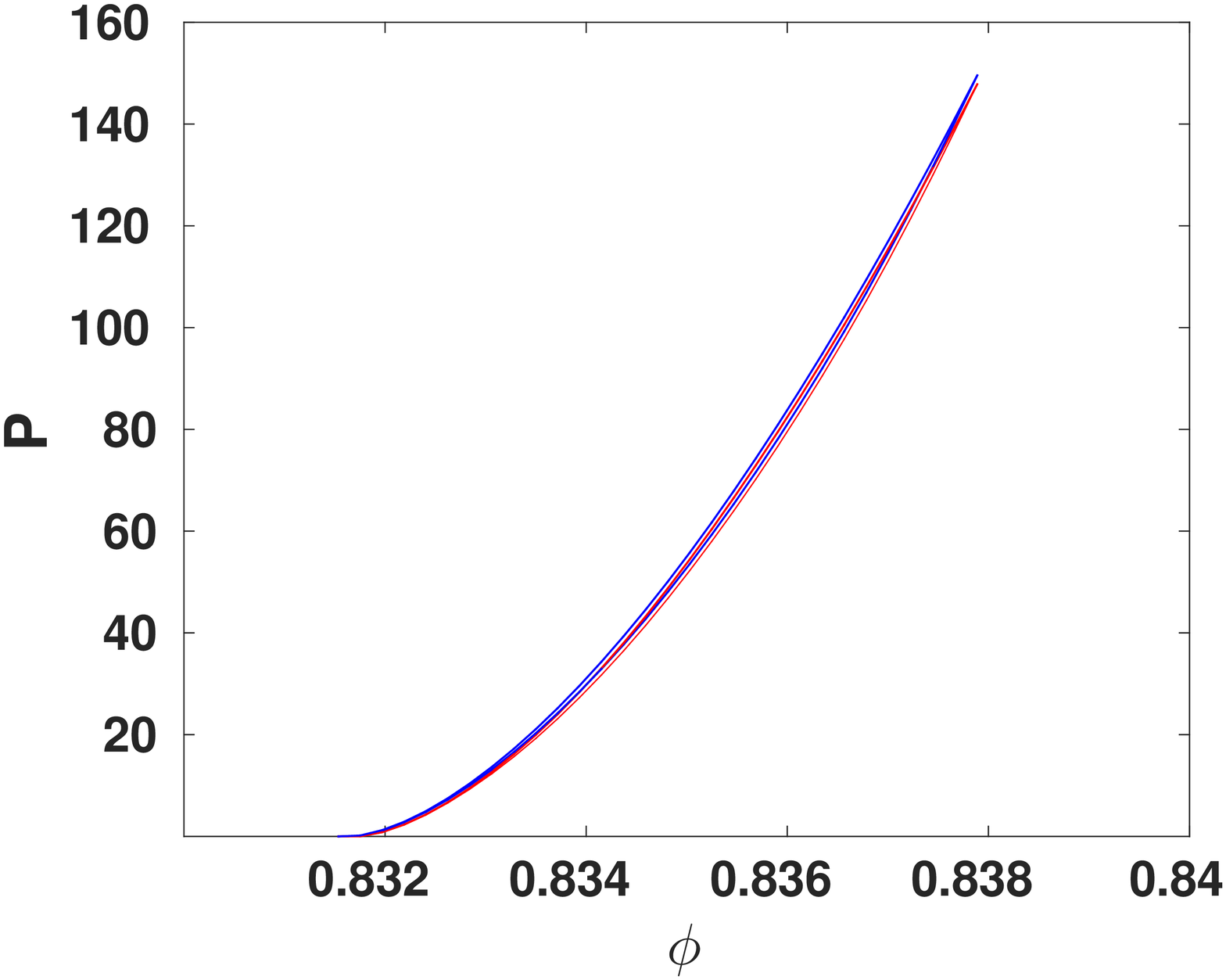}
\caption{Upper panel: examples of low order hysteresis loops in the $P-\Phi$ plane. The compression legs are blue and the decompression legs red. Lower panel: examples of higher order hysteresis loops in the $P-\Phi$ plane with the same color convention.  The high order loops are no longer able to compactify the system further, and
the compression leg begins at the same volume fraction where the decompression leg ends.}
\label{examples}
\end{figure}

In conclusion, we presented and explained a universal scaling law in the context of the cyclic training of an amorphous assembly of frictional disks. The protocol exhibits  a reduction in the dissipation per loop until the system reaches an asymptotic configuration with maximal volume fraction (for the maximal pressure chosen in the cyclic protocol).
Once achieved, the system has a perfect memory of the stressed state even when it is completely decompressed to zero pressure. Interestingly enough, repeated compressions are not without dissipation, since shear always induces frictional slips.
But the beginning and final volume fractions become invariant and the system repeats exactly the same hysteresis loop
in the $P-\Phi$ space. The amount of dissipation in the cyclic loops is governed by the scaling law Eq.~(\ref{plaw}) which
has a universal form with material dependent coefficients. The two terms in this equation were identified and related
to the dissipation due to changes in the neighbor list and the frictional slips respectively.

\acknowledgments{This work had been supported in part by the US-Israel BSF and the ISF joint grant with Singapore. IP is
indebted to the Niels Bohr International Academy for their hospitality under the Simons Fellowship where this paper was
written. MMB and the experiments were supported by the Collective Interactions Unit, OIST Graduate University.}

\clearpage

\title{Supplemental Material: Training, Memory and Universal Scaling in Amorphous Frictional Granular Matter}

\author{M. M. Bandi$^1$}
\author{H. George E. Hentschel$^2$}
\author{ Itamar Procaccia$^{2,3}$}
\author{Saikat Roy$^2$}
\author{Jacques Zylberg$^2$}
\affiliation{$^1$ Collective Interactions Unit, OIST Graduate University, Onna, Okinawa, 904-0495 Japan. \\
$^2$Dept of Chemical Physics, The Weizmann Institute of Science, Rehovot 76100, Israel.\\$^3$The Niels Bohr International Academy, University of Copenhagen, Blegdamsvej 17, DK-2100 Copenhagen, Denmark}

\maketitle

 \section{Details of the Experimental System}
 \subsection{Experimental Design}

The basic experimental design (see Fig.~2 upper panel of main text for schematic) follows the setup reported in Ref. \cite{Bandi2013} with several design improvements to be elaborated below. It consisted of a quasi two-dimensional chamber constructed from a steel frame of inner dimensions 0.6 m $\times$ 1.1 m $\times$ 0.02 m, with a transparent acrylic bottom plate.  A transparent Teflon sheet was glued to the top side of the acrylic bottom plate facing into the chamber to reduce friction between disks and the bottom plate. Whereas our primary measurements are conducted with disks comprised of stress birefringent (photoelastic) polymer, the current experiments were conducted with disks machined out of Lexan Polycarbonate. We do not detail photoelastic measurements here. The top and bottom of each disk was glued with a transparent teflon sheet, but the sides were intentionally machined with high roughness to achieve inter-disk contact friction responsible for friction-induced hysteresis studied here. The machined roughness yielded a static friction coefficient of $\mu = 0.21$ between disk contacts as measured using the method detailed in Ref. \cite{Bandi2013}. The friction between disks and bottom plate was measured to have a static friction coefficient of $\mu = 0.04$.

\begin{figure*}
\begin{center}
\includegraphics[width = 7 in]{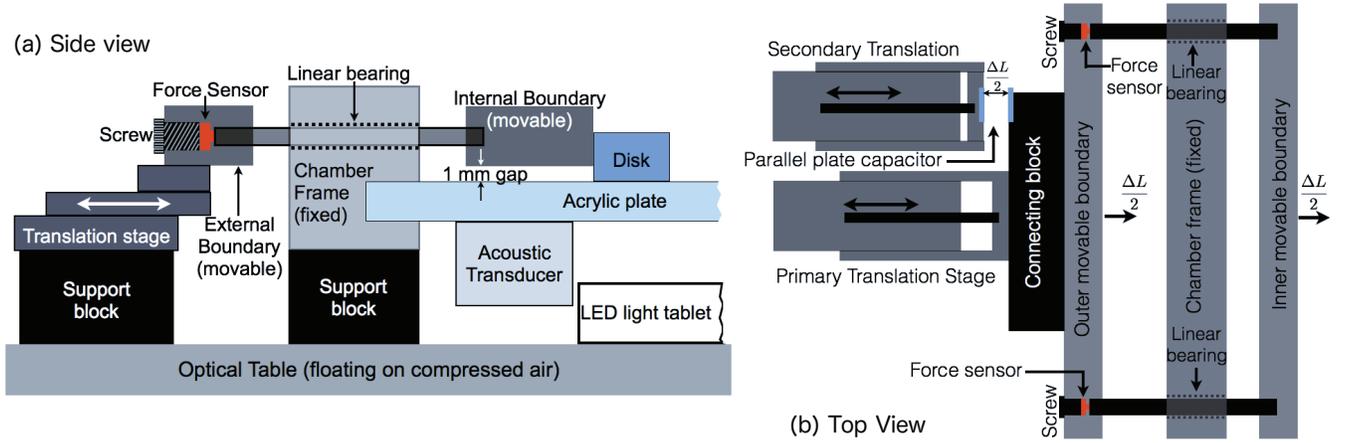}
\end{center}
\caption{(Color online) Compression axis schematic (a) Side view: A linear bearing passes through the interrogation chamber frame and connects an internal boundary to an external one. The internal boundary avoids frictional contact with the acrylic bottom (1 mm gap), but makes contact with, and moves disks. A force sensor (red) located within the external boundary makes contact with the linear bearing rod and is held rigid in place by a screw. A circular polarized DC LED light source illuminates the pack from below. (b) Top view: The primary translation stage moves the outer and inner boundaries. A secondary translation stage is capacitively coupled to the movable boundary through a parallel plate capacitor (light blue). Both primary and secondary stages are motorized by LabView. After adjusting initial capacitor gap to 50 nm, the secondary stage is held stationary while the primary stage displaces boundaries and increases the capacitor gap. A 5 V DC signal across the capacitor is sampled at 10 KHz, constantly monitors displacement and achieves a minimum precise quasi-static step of $\Delta L/2 = 250$ nm per boundary through closed loop LabView control.}
\label{MemSIFig1}
\end{figure*}

Our experiments were designed to achieve high precision in translation using feedback control {\it via} capacitive displacement sensors. The setup therefore demanded very high machining tolerance in dimensions of the chamber frame as well as the internal and external movable and immovable boundaries. Since it is impossible to achieve machining tolerance of 100 nm over part lengths of meter dimensions, the frame and boundary parts were machined in individual  pieces of 10 cm length on Computer Numerically Controlled mills. These individual parts were then assembled in an interlocking French cleat mechanism with a 37.5$^{\circ}$ ratchet geometry and held by screws to assure structural rigidity to obtain a contiguous structure. The assembled chamber was clamped rigid to an optical table standing on graded concrete foundation and floated by compressed air to isolate extraneous ground vibrations. A bidisperse set of large (diameter $D_L = 1.5$ cm) and small (diameter $D_S = 1$ cm) disks of thickness 0.975 cm in equal number ratio, were placed in the quasi two-dimensional experimental chamber with two opposing movable boundaries (Compression Axis) and two transverse, immovable boundaries (Transverse Axis). Four acoustic transducers placed asymmetrically as shown in Fig.~2 upper panel of main text (also see Fig.~\ref{MemSIFig1}a), provided spatially homogeneous acoustic excitation to the pack as explained later.

An internal steel boundary extended 5 cm into the chamber from each chamber wall, thus providing actual interrogation chamber dimensions of 0.5 m ($L$) $\times$ 1 m ($W$) $\times$ 0.02 m ($D$). The internal boundaries are what we show in the schematic presented in Fig.~2 upper panel of main text marked ``Movable'' and ``Fixed'', and in fig.~\ref{MemSIFig1}a and b as the ``Internal movable boundary'' for the compression axis perimeter. As schematically shown in fig.~\ref{MemSIFig1}a, the internal boundaries maintain a 1 mm vertical gap with the bottom acrylic plate, but do contact disks whose height extends to 0.975 cm. Four steel rods connect each internal boundary to an external boundary by passing through high precision linear bearings placed within the steel chamber frame. A precision force sensor was located at the terminating point of each steel rod within the external boundary to measure boundary pressure of the granular pack. A total of 16 sensors (4 per boundary) provided the boundary pressure read out. In the absence of frictional contact between internal boundaries and acrylic bottom, and high lubrication translation provided by linear bearings, the forces measured by the sensors were predominantly those experienced by the granular pack alone, with no systematic errors in measured signals as will be explained in the following.

The external boundaries along the transverse axis were clamped rigid to the floating optical table thereby rendering the transverse axis boundaries immovable. On the other hand, the external boundaries along the compression axis connected to a high precision motorized translation stage to achieve quasi-static compression. A circular DC light source illuminated with cold LEDs was placed under the experimental chamber to provide backlit illumination through the transparent acrylic bottom. Since measurements reported here only concern boundary pressure measurements, and do not involve imaging, the details will be presented elsewhere.

\begin{figure*}
\begin{center}
\includegraphics[width = 7 in]{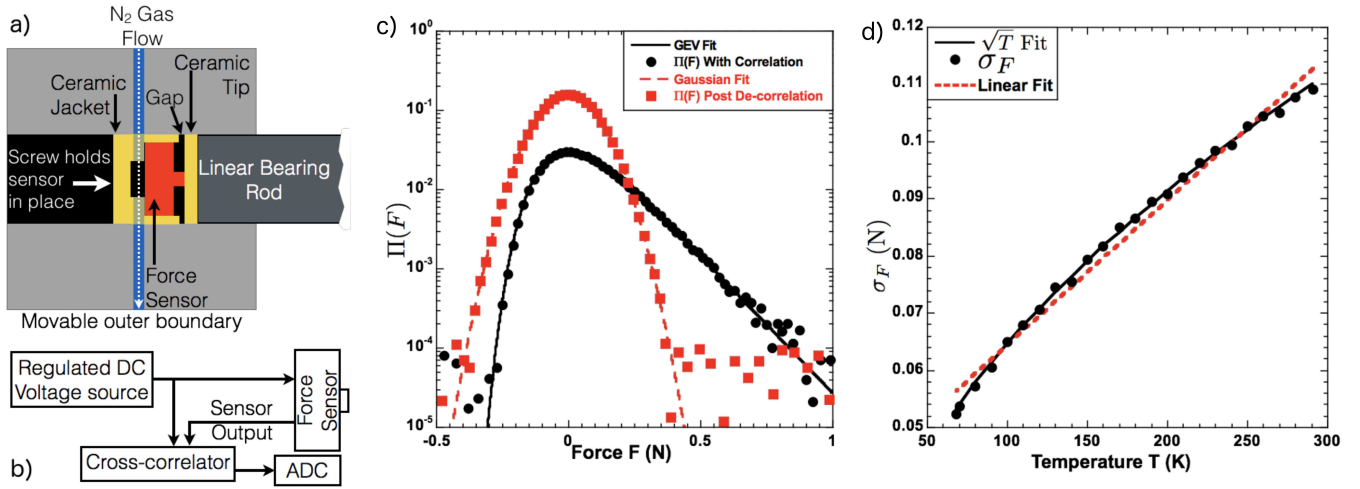}
\end{center}
\caption{(Color online) (a) The force sensor (red) within the external boundary is enclosed in a heat insulating ceramic jacket (yellow). Its nipple makes contact with the linear bearing rod with a ceramic heat insulating disk glued to its end. Nitrogen (N$_2$ gas) at 68K flows through to cool the sensor. (b) Electronic circuit schematic for correlated noise removal has two inputs, one from the DC regulated voltage source which drives the sensor and the sensor output. Noise in sensor output due to voltage driver is spectrally filtered out and fed to LabView data acquisition system. (c) Probability density function (PDF) of force with correlated noise (solid black circle) follows the Generalized Extreme Value (GEV) distribution (solid black line) whereas the noise post cross-correlator stage (solid red squares) is fit very well by a Gaussian (dashed red line). Force fluctuations measured post cross-correlation follows Johnson-Nyquist form and (d) its standard deviation $\sigma_F$ (solid black circle) falls as square root of Temperature (solid black line). The linear fit (dashed red line) is plotted for comparison.}
\label{MemSIFig2}
\end{figure*}

\subsection{Movable Boundary Translation}
The compression axis boundary positions were controlled by very high precision motorized translation stages as schematically shown in fig.~\ref{MemSIFig1}a and b. A Newport CONEX MFA-CC servo translation stage with 250 mm translational distance and 100 nm minimum step size was employed for automated translation control. We found the CONEX-CC controller usually supplied with the stage did not offer the precision we desired. We therefore constructed a precision displacement sensor in house to measure quasi-static compressive displacements along the movable boundary. Our choice of inhouse design for the displacement sensor was dictated by experimental considerations. We found optical interferometric displacement sensors failed due to unavoidable mechanical vibrations induced during the experimental protocol. Instead we employed a set of two translation stages separated by 25 nm, and a homebuilt parallel plate capacitor was installed as shown in fig.~\ref{MemSIFig1}b. The parallel place capacitor was used as a capacitive displacement sensor and was interfaced with a LabView data acquisition and control system to monitor displacements at 10 KHz sampling frequency and control compression axis quasi-static displacements down to steps of $\Delta L/2 = 250$ nm. Our tests showed the real achievable displacement resolution was 17.6 nm with an uncontrollable backlash introduced through the stage manufacturing process of 2.35 nm, both being far below our desired displacement resolution of 250 nm.

We employed the following quasi-static compression protocol:\\
1) At the start of each quasi-static step, the secondary stage was moved into position to achieve a parallel plate capacitor gap of 50 nm between the primary and secondary stages.\\
2) Keeping the secondary stage position fixed, the primary stage was translated at a speed of 1 mm/s, with capacitance being monitored at 10 KHz sampling frequency by the LabView system.\\
3) The closed loop control circuit implemented in LabView automatically stopped the primary translation stage once a distance of $\Delta L/2$ was achieved, where the minimum achievable $\Delta L/2$ was 250 nm.\\
4) This quasi-static compressive displacement transforms to a quasi-static step in packing fraction $\Delta \Phi = V/(\Delta L \times W)$, since both opposing movable boundaries were translated simultaneously. This latter point albeit subtle, becomes important in light of the strong protocol dependence observed in frictional measurements.

\subsection{Boundary Force Measurement}
The boundary pressure was measured with subminiature force sensors (LCMKD-50N, Omegadyne). Each force sensor was placed within the external boundary and held in place by a screw from one end. At the other end, the nipple of the button sensor made contact during compression with the linear bearing rod which terminated within the external boundary. The sensor was encased within a 3D printed (Visijet PXL-Core 3D printer) heat insulating ceramic jacket and a ceramic tip was glued to the teminating end of the linear bearing rod as shown in Fig.~\ref{MemSIFig2}a. A small gap was maintained between the sensor and linear bearing rod through which Nitrogen gas at 68K flowed through a hole drilled in the external boundary to cool down the force sensor.

The raw force sensor output exhibits correlated noise from several sources, including ground loops, capacitive coupling, and correlated noise from the regulated DV voltage circuit that drives the force sensor. After accounting for ground loops and capacitive coupling in the system, a noise cross-correlator circuit designed inhouse was employed (Fig.~\ref{MemSIFig2}b). Output from the regulated DC voltage circuit was split with one line driving the force sensor and the other forming one of two inputs to the cross-correlator circuit. The force sensor output formed the second input to the cross-correlator, which spectrally filtered the noise in force output arising from fluctuations in the DC regulated voltage supply. The output from the cross-correlator was then sent to a LabView data acquisition system for Analog to Digital Conversion (ADC).

Figure~\ref{MemSIFig2}c shows the probability density function (PDF) of force sensor output with (solid black circles) correlated noise, i.e. prior to entering cross-correlator circuit, and is fit very well by the Generalized Extreme Value (GEV) or the Fischer-Tippett distribution (solid black line in Fig.~\ref{MemSIFig2}c) of the form:
\begin{equation}
\Pi(F) = \frac{1}{\sigma_F}t(F)^{\xi + 1}e^{-t(F)}
\end{equation}
where $t(F) = \left(1+\left(\frac{F - \langle F \rangle}{\sigma_F}\right)\xi \right)^{-1/\xi}$ if $\xi \neq 0$ and $t(F) = e^{-(F - \langle F \rangle )/\sigma_F}$ if $\xi = 0$. Here $F$ is measured force in Newtons, $\langle F \rangle$ is mean force time-averaged over the duration of signal acquisition, $\sigma_F$ is the standard deviation, with $\xi$ forming the only fit parameter for the data, which was found to be $\xi = 0.5 \simeq 0$, and accordingly $t(F) = e^{-(F - \langle F \rangle )/\sigma_F}$.

On the other hand, the force sensor output post cross-correlator stage (solid red squares in Fig.~\ref{MemSIFig2}c) exhibits nearly Gaussian fluctuations (dashed red curve in Fig.~\ref{MemSIFig2}c). This classical Johnson-Nyquist form for post cross-correlator noise permits one to employ noise reduction by exploiting the fluctuation-dissipation theorem by virtue of the fact that the force sensor output is a voltage which is linearly proportional to the measured force. Just as Johnson-Nyquist noise follows the form $V_{rms} = \sqrt{4k_BTR\Delta f}$ where $k_B$ is Boltzmann constant, $T$ is temperature, $R$ is resistance and $\Delta f$ is the frequency bandwidth, we have for the force measurement:
\begin{equation}
\sigma_F  \propto \sqrt{4k_BTR\Delta f}
\end{equation}
Indeed, cryogenically cooling the electronics and force sensors exhibits noise reduction with a square-root dependence on temperature $T$ as shown in Fig.~\ref{MemSIFig2}d. There, the standard deviation in force $\sigma_F$ (solid black circles) fell as we cooled the circuits from room temperature ($T = 291$ K) down to $T = 68$ K. The square-root fit with temperature (solid black line) is decidedly better than a linear fit (dashed red line). We found flicker noise (Shot Noise) if the circuits were cooled below $T = 68$ K. Rather than employ further electronic measures, we instead adopted noise averaging by sampling the force at 1 KHz for 10 second duration and taking its average. Since the uncorrelated noise is expected to average as $1/\sqrt{N}$ where $N$ is the number of force measurement samples, one achieves a further noise reduction. The final precision we achieved in force measurement was 5.3 mN.

\subsection{Acoustic Perturbation}
Mechanical perturbation of granular media is usually achieved by applying vibrations at the boundaries, but such schemes are not spatially homogeneous because the dissipative collisions among particles cause a gradient in the perturbation magnitude as one moves from the boundary into the system interior. Spatially homogenous perturbation being necessary for the current experiments, we stuck four acoustic transducers (SD1G from Solid Drive) to the bottom of the acrylic plate (see Fig.~\ref{MemSIFig1}a) with an asymmetric placement as shown in Fig.~2 of main text. With acoustic energy being transferred to the acrylic bottom plate, it acted as the speaker and thus provided homogeneous perturbation. Each transducer was powered by an amplifier (SD250 from Solid Drive) connected to an independent function generator which output white noise with a frequency cutoff of 15 KHz, thereby providing a reasonable approximation for $\delta$-correlation in time. However, acoustic waves have long correlation lengths in the transmitted medium as one trivially observes in Chaldni patterns; achieving a reasonable approximation for $\delta$-correlation in space is therefore not so straightforward. In a beautiful study, Cadot et al demonstrated \cite{Cadot2010} nonlinear response when an elastic medium is subjected to random acoustic forcing. In addition to approximately Gaussian in time acoustic excitation, we additionally scrambled the amplitudes of all four transducers with a fifth function generator that provided band-limited (to 15 KHz) white noise. The four transducer amplitudes were scrambled in a manner such that their sum was always constant at any given instant.

\begin{figure}
\begin{center}
\includegraphics[width = 3.5 in]{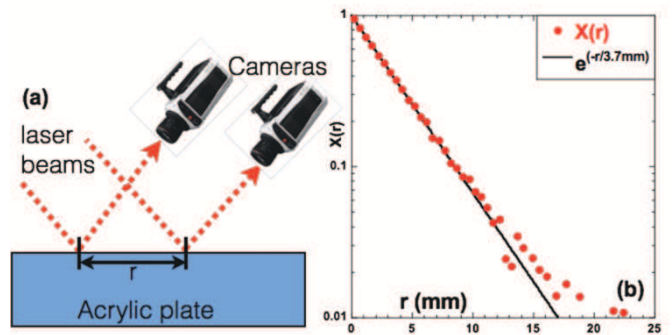}
\end{center}
\caption{(Color online) (a) Laser vibrometry schematic: Two laser beams incident onto the bottom acrylic plate under acoustic excitation at two locations separated by distance $r$. Height variations caused by acoustic waves cause perturbations in the reflected laser beams. These perturbations are captured by two high-speed cameras (Phantom v640) at 1500 frames per second, and the images are analysed to obtain the instantaneous spatial cross-correlation $X(r)$. (b) $X(r)$ (solid red circles) versus $r$ in log-linear scale shows exponential decay with a decay length of 3.7 mm obtained from data fit (solid black line).}
\label{MemSIFig3}
\end{figure}

Following the protocol of Cadot et al, we used laser vibrometry (see Fig.~\ref{MemSIFig3}a for schematic) to measure spatial cross-correlations in surface deformations. The cross-correlation function is defined as:
\begin{equation}
X(r) = \frac{\langle h(\vec{r_1},t) h(\vec{r_2},t) \rangle}{\sigma_{h(\vec{r_1})} \sigma_{h(\vec{r_2})}}
\end{equation}
where $h(\vec{r_1},t)$ and $h(\vec{r_2},t)$ are instantaneous (at time $t$) height variations at positions $\vec{r_1}$ and $\vec{r_2}$ respectively, and $\sigma_{h(\vec{r_1})}$ and $\sigma_{h(\vec{r_2})}$ are standard deviations of heights at positions $\vec{r_1}$ and $\vec{r_2}$ respectively, where the height variation $h$ was recorded by high-speed cameras (Phantom v640) that captured laser beams reflected off the acrylic plate. This cross-correlation function $X(r)$ (solid red circles) is plotted versus distance $r = |\vec{r_1} - \vec{r_2}|$ in mm in log-linear scale in Fig.~\ref{MemSIFig3}. $X(r)$ exhibits an exponential decay with decay length of 3.7 mm obtained from fit to data (solid black line in Fig.~\ref{MemSIFig3}). This decay length of 3.7 mm being less than small disk diameter of 1 cm, we treat this as approximately $\delta$-correlated perturbation in space.

\section{Details of the Numerical Simulations}
Conjugate gradient method is generally used to study the frictionless granular materials \cite{97MLGLBW}; but when the particles have friction, MD simulation is preferred as it correctly keeps track of both normal and history dependent tangential forces. So we employ MD simulation for the current study. Simulation of uniaxial compression of two dimensional granular packings are performed using open source codes, LAMMPS \cite{95P} and LIGGGHTS \cite{12KGHAP}. To simulate the experimental system the particles are taken as bi-disperses disks of unit mass with diameters 1 and $1.4$ respectively. The particles are placed randomly in a three dimensional box of dimension, $57$ (along $x$), $102$ (along $y$) and $1.4$ (along $z$). Quasistatic compression is implemented by displacing the boundary particles. A side wall made of particles is placed in the direction perpendicular to the compression direction.

The contact fores (both the normal and tangential forces which arises due to friction) are modeled according to the DEM (discrete element method) developed by Cundall and Strack \cite{79CS}. Implementation of static friction is done via tracking the elastic part of the shear displacement from the time contact was first formed.
When the disks are compressed they interact via both
normal and tangential forces.  Particles $i$ and $j$, at positions ${\B r_i, \B r_j}$ with velocities ${\B v_i, \B v_j}$ and angular velocities ${\B \omega_i, \B \omega_j}$ will  experience a relative normal compression on contact given by $\Delta_{ij}=|\B r_{ij}-D_{ij}|$, where $\B r_{ij}$ is the vector joining the centers of mass and $D_{ij}=R_i+R_j$; this gives rise to a  normal force $ \B F^{(n)}_{ij} $. The normal force is modeled as a Hertzian contact, whereas the tangential force is given by a Mindlin force \cite{79CS}. Defining $R_{ij}^{-1}\equiv R_i^{-1}+R_j^{-1}$, the force magnitudes are,
\begin{eqnarray}
\B F^{(n)}_{ij}\!&=&\!k_n\Delta_{ij} \B n_{ij}-\frac{\gamma_n}{2} \B {v}_{n_{ij}}\ , \:
\B F^{(t)}_{ij}\!=\!-k_t \B t_{ij}-\frac{\gamma_t}{2} \B {v}_{t_{ij}} \\
k_n &=& k_n^{'}\sqrt{ \Delta_{ij} R_{ij}} \ , \quad
k_t = k_t^{'} \sqrt{ \Delta_{ij} R_{ij}} \\
\gamma_{n} &=& \gamma_{n}^{'}  \sqrt{ \Delta_{ij} R_{ij}}\ , \quad
\gamma_{t} = \gamma_{t}^{'}  \sqrt{ \Delta_{ij} R_{ij}} \ .
\end{eqnarray}
Here $\delta _{ij}$ and $t_{ij}$ are normal and tangential displacement; $\B n_{ij}$ is the normal unit vector.  $k_n^{'}$ and $k_t^{'}$ are spring stiffness for normal and tangential mode of deformation: $\gamma_n^{'}$ and $\gamma_t^{'}$ are viscoelastic damping constant for normal and tangential deformation.
   $\B {v_n}_{ij}$ and $\B {v_t}_{ij}$ are respectively normal and tangential component of the relative velocity between two particles. The relative normal and tangential velocity are given by
   \begin{eqnarray}
\B {v}_{n_{ij}}&=& (\B {v}_{ij} .\B n_{ij})\B n_{ij}  \\
\B {v}_{t_{ij}}&=& \B {v}_{ij}-\B {v}_{n_{ij}} - \frac{1}{2}(\B \omega_i + \B \omega_j)\times \B r_{ij}.
\end{eqnarray}
   where $\B {v}_{ij} = \B {v}_{i} - \B {v}_{j}$. Elastic tangential displacement $ \B t_{ij}$ is set to zero when the contact is first made and is calculated using $\frac{d \B t_{ij}}{d t}= \B {v}_{t_{ij}}$ and also the rigid body rotation around the contact point is accounted for to ensure that $ \B t_{ij}$ always remains in the local tangent plane of the contact \cite{01SEGHLP}.

   The translational and rotational acceleration of particles are calculated from Newton's second law; total forces and torques on particle $i$ are given by

      \begin{eqnarray}
\B F^{(tot)}_{i}&=& \sum_{j}\B F^{(n)}_{ij} + \B F^{(t)}_{ij}  \\
\B \tau ^{(tot)}_{i}&=& -\frac{1}{2}\sum_{j}\B r^{ij} \times \B F^{(t)}_{ij}.
\end{eqnarray}

   The tangential force varies linearly with the relative tangential displacement at the contact point as long as the tangential
   force does not exceed the limit set by the Coulomb limit
   \begin{equation}
   F^{(t)}_{ij} \le \mu F^{(n)}_{ij} \ , \label{Coulomb}
   \end{equation}
  where $\mu$ is a material dependent coefficient. When this limit is exceeded the contact slips in a dissipative
  fashion. In our simulations we reset the
  value of $t_{ij}$  so that $F^{(t)}_{ij} =0.8 \mu F^{(n)}_{ij}$. This choice is somewhat arbitrary, but recommended on the basis of frictional slip events measured in
  experiments in the laboratory of J. Fineberg \cite{Fine}. A global damping is implemented to reach the static equilibrium in reasonable amount of time. After each compression step, a relaxation step is added so that the system reaches the static equilibrium and then the global stress tensor is measured by taking averages of the dyadic products between the contact forces and the branch vector over all the contacts in a given volume,
  \begin{equation}
\sigma_{\alpha \beta} =\frac{1}{V}\sum_{j\neq i}\frac{r^{\alpha}_{ij} F^{\alpha}_{ij} }{2}
  \end{equation}

\begin{figure}
\includegraphics[scale=0.25]{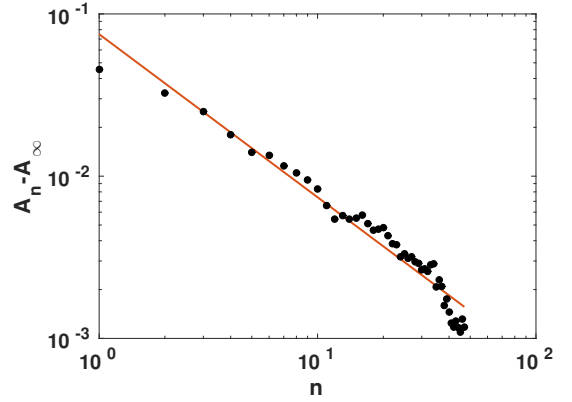}
\caption{The power law for the decaying areas under the hysteresis loops as measured in the numerical simulation. Here $\mu=0.3$,  Black dots are data and the red line is the best fitting power law with $\theta=-1.005$. }
\label{powerhighmu}
\end{figure}

\begin{figure}
\includegraphics[scale=0.35]{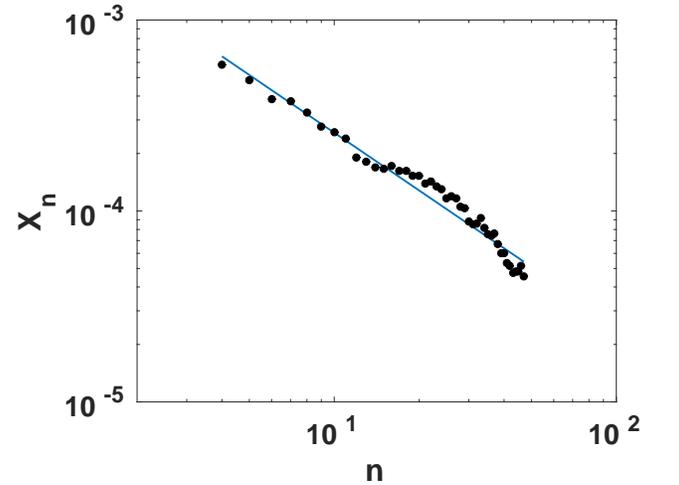}
\caption{Log-log plot of $X_n$ vs. $n$. Here $\mu=0.3$, the black dots are the data, the blue line is the best fitting scaling law with the exponents being -1.005.}
\label{Xnvsnhighmu}
\end{figure}

The pressure is determined from the trace of the stress and it is measured as a function of the packing fraction. After a full compression cycle, the packing is again decompressed to zero pressure and then again the next compression cycle begins. The area between the compression and decompression curve is also calculated as a function of number of cycles. We also calculate the total energy loss due to sliding events in each cycle. The system size is $N= 4000$ and the data are averaged over ten different initial configurations. We used two different friction coefficient $ 0.1$ and $0.3$.
 In Fig.~\ref{powerhighmu} we plot loop area after subtracting it from asymptote area as a function of cycles for friction coefficient of $0.3$ and the exponent of the power law remains the same as that of low friction case (See Fig. 4 main paper). We also plot $X_n$ as a function of number of cycles for the same friction coefficient  in Fig. ~\ref{Xnvsnhighmu} and the measured exponent (See Fig. 5 main paper) indicates the universality in the scaling law.

\end{document}